\tolerance=2500
\headline={\ifnum\pageno=1 \hfil
                      \else                         \hss\tenrm
Collier, Roberts, and Figueroa-Vi\~nas: Acoustic Kappa Waves
\hss\tenrm\folio\voffset=1.66in\fi}
\footline={\ifnum\pageno=1 \hss\tenrm\folio\hss
                    \else                          \hfil\fi}
\vskip 3.0in
\hoffset=-0.25in
\voffset=0.25in
\vsize=9.50in
\hsize=6.75in
\tenrm
\vskip 0.25in
\noindent
{\tenbf Acoustic Kappa-Density Fluctuation Waves in Suprathermal 
Kappa Function Fluids}
\vskip .10in
\noindent
\vskip 10pt \noindent
Michael R. Collier, Aaron Roberts, and Adolfo Vi\~nas 
\vskip 0.0in \noindent
{\it NASA/Goddard Space Flight Center, Code 673, Greenbelt, Maryland, 20771, USA}
\vskip 0.0in \noindent
{\it e-mails: Michael.R.Collier@nasa.gov, Aaron.Roberts@nasa.gov, Adolfo.Figueroa-Vinas-1@nasa.gov}
\vskip 10pt
\noindent
\vskip 0pt \noindent
date: 20 July 2007 -- version: 4.0 
\vskip 0pt \noindent
D3.1 Multiscale Magnetospheric Processes: Theory, Simulation, and Multipoint Observations
\vskip 0pt \noindent
COSPAR Paper Number: D3.1-0099-06
\vskip 0pt \noindent
keywords: kappa functions, suprathermal distributions, acoustic waves
\vskip 0 pt \noindent
\lineskiplimit=0in
\baselineskip=18pt
\lineskip=0in
\vskip .10in
{\noindent
\centerline{\tenbf Abstract}
\vskip .10in 
We describe a new wave mode similar to the acoustic wave in which both density and velocity fluctuate. Unlike the acoustic wave in which the underlying distribution is Maxwellian, this new wave mode occurs when the underlying distribution is a suprathermal kappa function and involves fluctuations in the power law index, $\kappa$. This wave mode always propagates faster than the acoustic wave with an equivalent effective temperature and becomes the acoustic wave in the Maxwellian limit as $\kappa\rightarrow\infty$.

%
%
%
\baselineskip=20pt
\vskip 20pt
\noindent{\tenbf 1. Introduction}

\nobreak

\vskip 5pt

\nobreak

Distributions in space plasmas are generally described by a drifting Maxwellian, 
$$\eqalignno{ f(x,v) &= {n(x)\over{\omega_0 \sqrt{\pi}}} \exp\{{-{(v-u)^2\over\omega_0^2}}\}, &(1)}$$
in one-dimension.
This type of distribution supports wave modes in which the various physical parameters: $n$, the density, $\omega_0$, the thermal speed, and $u$, the convection speed, can all oscillate. Perhaps the simplest of these is the acoustic or sound wave. 

However, space plasma distribution functions are not Maxwellian
[e.g. references in Collier, 1993], but suprathermal,
generally well-described by a drifting kappa function which in one dimension is [Summers and Thorne, 1991]:
$$\eqalignno{ f(x,v) &= {{n(x) \Gamma(\kappa+1)}\over{\omega_0 \sqrt{\pi}
\kappa^{3/2} \Gamma(\kappa-{1\over2})}}\,\,\,\, 
{1\over{{[1 + {(v-u)}^2/{\kappa \omega_0^2}]}^{\kappa}}}. &(2)}$$
Here, the parameters $n$ and $u$ are identical to the above, $\omega_0^2$ is a parameter analogous to that in equation~(1), and $\kappa$ is an additional parameter describing the prominence of the suprathermal tail, the lower the value of $\kappa$, the more pronounced the tail. In the limit as $\kappa\rightarrow\infty$, equation~(2) becomes equation~(1).

Although many view (2) as a non-equilibrium distribution, nature contradicts this notion. Suprathermal distributions well-described by kappa functions occur almost universally in space plasmas [e.g. Summers and Thorne, 1991; Collier, 1999] and theoretical considerations [Summers and Thorne, 1991; Leubner and V\"or\"os, 2005; Leubner, 2004a; Treumann et al., 2004], for example based on the Tsallis entropy [Tsallis, 1995], have aided in understanding their preeminence. In this manuscript, we shall adopt the point-of-view that the kappa function, rather than the Maxwellian, is the equilibrium distribution. Of course, we recover the Maxwellian results when $\kappa\rightarrow\infty$.

We have come to expect wave modes to involve the parameters of physical 
relevance identifiable in the Maxwellian distribution, namely the moments 
density, temperature (or equivalently $\omega_0$), and flow (or $u$).
However, there exists no physical reason why wave-like behavior 
should be limited solely to moments of the distribution. 
Because the kappa function, in addition to containing the associated 
aforementioned Maxwellian parameters, has the extra parameter $\kappa$, it seems reasonable, then, indeed expected, that for suprathermal distributions there should also be wave modes in which $\kappa$ participates. Here we provide an example of such a wave mode in a one-dimensional neutral fluid based on the ion acoustic wave.

Ion acoustic waves play an important role in the interaction between bodies moving supersonically through a plasma and the ambient plasma itself. In this case, they create a wake such as in the region behind Venus, the moon, Io and Titan. In the wake, converging beams of flowing plasma collide, exciting ion acoustic waves [e.g. Samir et al., 1983]. Broadband electrostatic noise that could plausibly be identified with ion acoustic wave activity has been observed in the polar cusp [Gurnett and Frank, 1978], the distant magnetotail [Gurnett et al., 1976] and along auroral field lines [Gurnett and Frank, 1977].

\vskip 20pt
\noindent{\tenbf 2. The Collision Operator}

\nobreak

\vskip 5pt

\nobreak
Although the presence of suprathermal tails may be interpreted as evidence of non-equilibrium distributions, the near-universal occurrence of these distributions in space plasmas [e.g. Summers and Thorne, 1991; Collier, 1999] argues in favor of a formalism which treats
these distributions as a physical equilibrium (but not thermodynamic 
equilibrium) state rather than as an anomaly.

To this end, we consider the one-dimensional force-free Boltzmann equation 
given by
$$\eqalignno{ {{\partial f}\over{\partial t}} +
v\cdot{{\partial f}\over{\partial x}}
&= {{{\partial f}\over{\partial t}}\vrule}_{\,c}\,, &(3)}$$
where $f$ is the distribution function, $v$ is the particle velocity,
$x$ is position, $t$ is time, and ${{{\partial f}/{\partial t}\,}\vrule}_{\,c}$
is the collision operator.

The standard approach is to assure that collisions reduce the distribution to a Maxwellian consistent with the local thermodynamic properties. However, in the case considered here, because the equilibrium distribution becomes the $\kappa$-function, the effect of the collision operator (under the appropriate constraints to be discussed shortly) must be consistent with this distribution.

Some properties of the collision operator remain unchanged in spite of the $\kappa$-function
assumption. Because collision operators are ``local" in configuration space, if two particles collide, they make a jump in velocity space but remain at approximately the same physical location before and after the collision.
Thus,
$$\eqalignno{\int_{-\infty}^{\infty}{ 
{{{\partial f}\over{\partial t}}\vrule}_{\,c}\, dv}
&= 0, &(4)}$$
because at a given position even though the particles there shift around in velocity space due to collisions, the local number density must not change.
Additionally, because momentum must be conserved in any collision
$$\eqalignno{\int_{-\infty}^{\infty}{ 
{{{\partial f}\over{\partial t}}\vrule}_{\,c}\, v\, dv}
&= 0. &(5)}$$

The standard treatment also assumes conservation of energy. However, there exists no reason to expect energy to be conserved in whatever process establishes the equilibrium kappa distribution. It can be shown that the assumption of a constant order of magnitude of energy $<\ln\{H\}>$, where $H = \Phi + {1\over2}mv^2$ with
$\Phi$ the (positive) potential, $m$ the mass, and $v$ the velocity, leads to an equilibrium distribution that is a kappa function [Collier, 2004]. Thus, the collision operator obeys
$$\eqalignno{\int_{-\infty}^{\infty}{ 
\ln\{\Phi + {1\over2}mv^2\} \, {{{\partial f}\over{\partial t}}\vrule}_{\,c}\, dv}
&= 0. &(6)}$$

One final note: this approach assumes that collisions are dominant and that the collision frequency is the highest frequency for the phenomenon under consideration. Of course, this is an invalid assumption for collisionless space plasmas. Nevertheless, we justify this assumption on the basis that this is the standard approach for deriving the fluid equations, which work far better than they should for space physics applications.

\vskip 20pt
\noindent{\tenbf 3. The Equation Set}

\nobreak

\vskip 5pt

\nobreak
The first two equations with which we will work are
$$\eqalignno{ {{\partial n}\over{\partial t}} + {{\partial}\over {\partial x}} 
[un] &= 0, &(7)\cr
nm{{du}\over{dt}} &= -{{\partial P}\over{\partial x}} =
-{{\partial}\over{\partial x}}(n {\kappa\over{\kappa-{3/2}}} 
\cdot {{1\over2} m\, \omega_0^2} ). &(8)}$$
Equations~(7) and (8) are simply the zeroth and first moments of equation~(3) which represent physically mass conservation and momentum. 
The time derivative on the left of equation~(8) is the convective derivative. 
In equation~(8), we have substituted for the pressure, $P$, the product of the density and temperature. Here we introduce an artificial ``core" temperature,
$T_{core}$, defined by
$$\eqalignno{ T &= {\kappa\over{\kappa-{3\over2}}} T_{core}
= {\kappa\over{\kappa-{3\over2}}}\cdot {{1\over2} m\, \omega_0^2}. &(9)}$$
The core temperature is simply the corresponding Maxwellian temperature for
$\kappa\rightarrow\infty$ and is not meant to imply that the distribution
has two temperatures since the temperature, a moment of the distribution
function, can be calculated uniquely for any distribution [Leubner, 2004b].
Note that the factor $\kappa-{3\over2}$ in the denominator appears in both the one-dimensional and three-dimensional cases.

The justification for defining the variable $T_{core}$ is that the expression for the entropy of the $\kappa$-function:
$$\eqalignno{ 
{{T_{core}^{1/2}\over n} \xi_{1D}(\kappa)} &= constant, &(10)}$$
where
$$\eqalignno{ 
{\xi_{1D}(\kappa)} &= { {\kappa^{3\over2} {\Gamma(\kappa-{1\over2})}}
\over {\Gamma({\kappa+1})} } \exp\biggl\{ {\kappa [\psi(\kappa) - \psi(\kappa-{1/2})] - {1/2}} \biggr\},
 &(11)}$$
and $\psi$ is the psi or digamma function equal to the derivative of the logarithm of the gamma function [Abramowitz and Stegun, 1972], will be used
to close the fluid equations. Because this expression involves separate factors depending only on $\omega_0$ and $\kappa$ respectively, defining the additional variable $T_{core}$ will prove convenient for the following discussion. The function $\xi_{1D}$ is shown in Figure~1 along with the three dimensional case for comparison [Collier, 1995].

Equation~(10) results from the $\ln\{f\}/n$ moment of equation~(3), noting that the equilibrium distribution is a kappa function, that is using equation~(6), and is physically conservation of entropy [Collier, 1995].
To achieve equation~(10), we consider $f(x,v,t) = n(x,t)\cdot g(x,v,t)$ where $g(x,v,t)$ is the normalized velocity space distribution, that is,
$$\eqalignno{\int_{-\infty}^{\infty} g(x,v,t) \, \, dv &= 1. &(12)}$$
Taking the $\ln\{n\,g\}/n$ moment of the right hand side of equation~(3),
the collision operator, gives us
$$\eqalignno{\int_{-\infty}^{\infty}  
{{{\partial f}\over{\partial t}}\vrule}_{\,c}
\cdot {1\over n} \ln\{n\,g\} \, dv &= 
{1\over n} \int_{-\infty}^{\infty} 
{{{\partial f}\over{\partial t}}\vrule}_{\,c} \ln\{g\} \, dv, &(13)}$$
because of conservation of particles as expressed by equation~(4).

The first term on the left hand side may be expanded as
$$\eqalignno{\int_{-\infty}^{\infty}  {{\partial f}\over{\partial t}}
\cdot {1\over n} \ln\{n\,g\} \, dv &= 
\int_{-\infty}^{\infty} {{\partial}\over{\partial t}} 
\bigl({{f\over n} \ln\{n\,g\}}\bigr) \, dv +
\int_{-\infty}^{\infty} \bigl({f\over n^2}\bigr) {{\partial n}\over{\partial t}}
\ln\{n\,g\} \, dv - 
\int_{-\infty}^{\infty} \bigl({f\over n^2}\bigr) 
{{\partial n}\over{\partial t}}\, dv, &(14)}$$
where we have used the fact that the integral of the time derivative of $g$ is zero because $g$ is normalized to unity, equation~(12).

Likewise, the second term on the left hand side may be expanded as
$$\eqalignno{\int_{-\infty}^{\infty}  v {{\partial f}\over{\partial x}}
\cdot {1\over n} \ln\{n\,g\} \, dv =
&\int_{-\infty}^{\infty} {{\partial}\over{\partial x}}
\bigl({{{f\,v}\over n} \ln\{n\,g\}}\bigr) \, dv +
\int_{-\infty}^{\infty} \bigl({{f\,v}\over n^2}\bigr) 
{{\partial n}\over{\partial x}} \ln\{n\,g\} \, dv -\cr
&\int_{-\infty}^{\infty} \bigl({{v\,f}\over n^2}\bigr) 
{{\partial n}\over{\partial x}}\, dv -
\int_{-\infty}^{\infty} v\, {{\partial g}\over{\partial x}}\, dv. &(15)}$$

In the sum of equations~(14) and (15), the last term in equation~(14) and the two last terms in equation~(15) combine to yield
$$\eqalignno{-{1\over n}\biggl[{{\partial n}\over{\partial t}} +
u\,{{\partial n}\over{\partial x}} +
n\,{{\partial u}\over{\partial x}}\biggr] &= 0, &(16)}$$
where the terms in the square brackets sum to zero by mass conservations, equation~(7).

Consequently, we can write the left hand side of the $\ln\{f\}/n$ moment of 
equation~(3)
$$\eqalignno{\int_{-\infty}^{\infty}  {{\partial f}\over{\partial t}}
\cdot {1\over n} \ln\{n\,g\} \, dv &+
\int_{-\infty}^{\infty}  v {{\partial f}\over{\partial x}}
\cdot {1\over n} \ln\{n\,g\} \, dv = \cr
\int_{-\infty}^{\infty} {{\partial}\over{\partial t}} 
\bigl({{f\over n} \ln\{n\,g\}}\bigr) \, dv +
\int_{-\infty}^{\infty} \bigl({f\over n^2}\bigr) {{\partial n}\over{\partial t}}
\ln\{n\,g\} \, dv &+
{{\partial}\over{\partial x}}\bigl[u\int_{-\infty}^{\infty} 
{{f\over n} \ln\{n\,g\}} \, dv \bigr] +
\int_{-\infty}^{\infty} \bigl({{f\,v}\over n^2}\bigr) 
{{\partial n}\over{\partial x}} \ln\{n\,g\} \, dv, &(17)}$$
where the third term on the right hand side arises from the assumption that $f$ is symmetric around $u$ so that
$$\eqalignno{\int_{-\infty}^{\infty}  \bigl( v-u \bigr)\,
{f\over n} \ln\{n\,g\} \, dv \,&=\, 0, &(18)}$$
which is equivalent to asserting that there is no heat flux.
Equation~(17), then, may be re-written as 
$$\eqalignno{\int_{-\infty}^{\infty}  {{\partial f}\over{\partial t}}
\cdot {1\over n} \ln\{n\,g\} \, dv &+
\int_{-\infty}^{\infty}  v {{\partial f}\over{\partial x}}
\cdot {1\over n} \ln\{n\,g\} \, dv = \cr
&{d\over{dt}} \bigl[ \int_{-\infty}^{\infty} 
{g \ln\{n\,g\}} \, dv \bigr] +
\int_{-\infty}^{\infty} {f\over n^2} \ln\{n\,g\} \,
\bigl[ {{\partial n}\over{\partial t}} + n {{\partial u}\over{\partial x}} +
u {{\partial n}\over{\partial x}} \bigr]\, dv, &(19)}$$
where $d/dt$ is the convective derivative, and we have again used equation~(18).

Noting that the terms in the square bracket in the second integral on the right hand side of equation~(19) sum to zero by mass conservation, equation~(16), 
we arrive at
$$\eqalignno{ {d\over{dt}} \bigl[ \int_{-\infty}^{\infty} 
{g \ln\{n\,g\}} \, dv \bigr] &=
{1\over n} \int_{-\infty}^{\infty} 
{{{\partial f}\over{\partial t}}\vrule}_{\,c} \ln\{g\} \, dv, &(20)}$$
where we have used the right hand side of equation~(13).

As discussed in Section~(2), we assume that the collision operator effects an equilibrium kappa distribution in accordance with Collier [2004] so that
$$\eqalignno{ g(v) &\propto {\bigl(\Phi + {1\over 2}mv^2\bigr)}^{-\kappa}. &(21)}$$
Under this condition, the right hand side of equation~(20) is zero, so that the 
integral
$$\eqalignno{ \int_{-\infty}^{\infty} {g \ln\{n\,g\}} \, dv &=
constant. &(22)}$$
Using the $g$ defined by the one-dimensional kappa function, equation~(2), noting that 
$$\eqalignno{ 
{{{\partial}\over{\partial \alpha}}\vrule}_{\,\alpha=0}
{\bigl(1 + {{v^2}\over{\kappa\omega_0^2}} \bigr)}^{\alpha}
&= \ln\bigl\{ {1 + {{v^2}\over{\kappa\omega_0^2}} } \bigr\}, &(23)}$$
using the identity
$$\eqalignno{ \beta(z,w) &= \int_0^\infty { {t^{z-1}}\over{{(1+t)}^{z+w}} }\, dt
= {{\Gamma(z) \Gamma(w)}\over {\Gamma(z+w)}}, &(24)}$$
and the definition of $\psi$
$$\eqalignno{ {{d\Gamma(z)}\over{dz}} &= \Gamma(z) \psi(z), &(25)}$$
one retrieves equation~(10) which is the entropy of the one-dimensional kappa
function. Note that using the large $z$ asymptotic expansions
$$\eqalignno{ \Gamma(z) &\approx \sqrt{2\pi} \exp\{-z\} z^{z-1/2}, &(26)\cr
\psi(z) &\approx \ln\{z\} - {1\over{2z}}, &(27)}$$
one can show that 
$$\eqalignno{ \lim_{\,\,\,\,\,\,\,\kappa\rightarrow\infty} {\xi_{1D}(\kappa)} &= 1, &(28)}$$
so that equation~(10) becomes the standard expression for conservation of entropy in a Maxwellian gas.

\vskip 20pt
\noindent{\tenbf 4. The Dispersion Relation: Standard-Acoustic Mode}

\nobreak

\vskip 5pt

\nobreak
Summarizing Section~3, the set of equations with which we will work are
$$\eqalignno{ {{\partial n}\over{\partial t}} + {{\partial}\over {\partial x}} 
[un] &= 0, &(29)\cr
nm{{du}\over{dt}} &= -{{\partial P}\over{\partial x}} =
-{{\partial}\over{\partial x}}(n {\kappa\over{\kappa-{3/2}}} T_{core}), &(30)\cr
{{T_{core}^{1/2}\over n} \xi_{1D}(\kappa)} &= const, &(31)}$$
and we will apply the standard approach of linearizing these equations assuming sinusoidal variation of the relevant parameters.

Equation~(29), mass conservation, requires that both density $n$ and velocity $u$ oscillate. So, we consider the standard equilibrium in which density oscillates around some equilibrium $n_0$ and ignore any equilibrium convection. However, {\it both $T_{core}$ and $\kappa$ need not oscillate simultaneously\/} so that this set of equations supports two distinct modes. We will consider first the mode in which core temperature $T_{core}$ oscillates but the suprathermal tail $\kappa$ does not so that
$$\eqalignno{ n &= n_0 + \delta n \exp\{i(kx-\omega t)\}, &(32)\cr
u &= \delta u \exp\{i(kx-\omega t)\}, &(33)\cr
T_{core} &= T_{core_0} + \delta T_{core} \exp\{i(kx-\omega t)\}, &(34)}$$
where the $\delta$'s correspond to first order perturbations.

Expanding equations~(29) through (31) to first order using equations~(32) through (34) gives us
$$\eqalignno{ \omega \delta n &= n_0 k \delta u, &(35)\cr
\omega \delta u &= {T_{core_0}\over m} {\kappa\over{\kappa-{3\over2}}} \bigg[ {{\delta n}\over n_0} + {{\delta T_{core}}\over{T_{core_0}}} \bigg] k, &(36)\cr
{{1\over2} {{\delta T_{core}}\over{T_{core_0}}}}&= {{\delta n}\over n_0}. &(37)}$$
Combining equations~(36) and (37) to eliminate ${{\delta T_{core}}/{T_{core_0}}}$
and then using equation~(35) to eliminate $\omega \delta u$ eventually gives us
$$\eqalignno{ \omega &= k \sqrt{ {{3 T}\over m} }, &(38)}$$
which is simply the dispersion relation for the one-dimensional acoustic wave [e.g. Chen, 1984] except that, in this case, the underlying distribution is a kappa function with suprathermal tails that simply ``go along for the ride," so to speak. Though, note that $T$ in equation~(38) above is the total temperature, not core temperature, so that the suprathermal character is still important in that sense.

\vskip 20pt
\noindent{\tenbf 5. The Dispersion Relation: Kappa-Acoustic Mode}

\nobreak

\vskip 5pt

\nobreak
Now we will make the assumption that only density, velocity and kappa oscillate according to
$$\eqalignno{ n &= n_0 + \delta n \exp\{i(kx-\omega t)\}, &(39)\cr
u &= \delta u \exp\{i(kx-\omega t)\}, &(40)\cr
\kappa &= \kappa_0 + \delta \kappa \exp\{i(kx-\omega t)\}, &(41)}$$
where the $\delta$s correspond to first order perturbations.

Using equations~(29), (30) and (31), we get to first order
$$\eqalignno{ \omega \delta n &= n_0 k \delta u. &(42)\cr
\omega \delta u &= {T_{core}\over m} {\kappa_0\over{\kappa_0-{3\over2}}} \bigg[ {{\delta n}\over n_0} - {{{3\over2}\delta\kappa}\over
{\kappa_0(\kappa_0-{3\over2})}}\bigg] k. &(43)\cr
{{\partial \ln\{\xi_{1D}\}}\over{\partial \kappa}} \delta\kappa &= {{\delta n}\over n_0}. &(44)}$$
Combining equations~(43) and (44), we get 
$$\eqalignno{ \omega \delta u &= {T_{core}\over m} {\kappa_0\over{\kappa_0-{3\over2}}} k \bigg[ {{\partial \ln\{\xi_{1D}\}}\over{\partial \kappa}} - {{3\over2}\over {\kappa_0(\kappa_0-{3\over2})}}\bigg] \delta\kappa. &(45)}$$
Combining equations~(42) and (44), we get
$$\eqalignno{ \omega \delta u &= {\omega^2\over k} {{\partial \ln\{\xi_{1D}\}}\over{\partial \kappa}} \delta\kappa. &(46)}$$
Finally, combining equations~(45) and (46), we arrive at the dispersion relation
for the one-dimensional kappa-acoustic wave:
$$\eqalignno{ \omega &= k \sqrt{ {3 T\over m} } \cdot
\sqrt{ {1\over 3} - {1\over2} \cdot {1\over{\kappa_0 (\kappa_0-{3\over2})
{{\partial \ln\{{\xi_{1D}(\kappa)}\}}\over{\partial \kappa}} }} }. &(47)}$$
Note that $\omega$ in equation~(47) is never imaginary.

\vskip 20pt
\noindent{\tenbf 6. Discussion}

\nobreak

\vskip 5pt

\nobreak
By examining equation~(47) and noting that the phase speed of a standard one-dimensional acoustic wave $c_s$ with a temperature $T$ is
$$\eqalignno{ c_s &= \sqrt{ {{3T}\over m} }, &(48)}$$
we can define the ratio of the phase speed of the kappa-acoustic wave
$v_\kappa$ to the phase speed of a standard acoustic wave of the same temperature as
$$\eqalignno{ {v_{\kappa}\over c_s} &= 
\sqrt{ {1\over 3} - {1\over2} \cdot {1\over{\kappa_0 (\kappa_0-{3\over2})
{{\partial \ln\{\xi_{1D}(\kappa)\}}\over{\partial \kappa}} }} }. &(49)}$$
By using the definition of $\xi_{1D}(\kappa)$ given by equation~(11) and
noting that $\psi(\kappa+1) - \psi(\kappa) = 1/\kappa$,
we can show that
$$\eqalignno{ {{\partial\ln\{\xi_{1D}(\kappa)\}}\over{\partial\kappa}} &= 
{1\over{2\kappa}} + \kappa[\psi^\prime(\kappa) - \psi^\prime(\kappa-{1\over2})]. &(50)}$$
Here $\psi^\prime$, the derivative of the gamma function, is referred to as the trigamma function [Abramowitz and Stegun, 1972].
Figure~(2) shows equation~(49) plotted as a function of kappa. The phase velocity of the acoustic kappa wave is always greater than a standard acoustic wave of the same temperature and the two phase velocities are identical in the limit of large $\kappa$, as can be shown by using the asymptotic expression for large $\kappa$,
$\psi^\prime(\kappa) \sim 1/\kappa + 1/2\kappa^2$.
The phase velocity of the kappa wave diverges to infinity as the value of kappa goes to $3/2$.

In addition to the obvious characteristic that in the kappa acoustic mode the suprathermal tail oscillates, note that for the kappa acoustic wave, 
the relative density fluctuation obeys 
$$\eqalignno{ 
{{\delta n}\over n_0} &= {{\partial \ln\{\xi_{1D}\}}\over{\partial \kappa}} \delta\kappa. &(51)}$$
while for the standard-acoustic mode the relationship is 
$$\eqalignno{ 
{{\delta n}\over n_0} &= {{1\over2} {{\delta T_{core}}\over{T_{core_0}}}}, &(52)}$$
so that in the kappa acoustic mode, the compressions in density are associated
with decreases in $\kappa$ or more pronounced tails. This, then, is the equivalent
of ``heating" the distribution except that the energy goes into the tail rather
than the core. However, the restoring force is different because the degree to
which the tail can be tweaked is different than that of the core and is given
by the equation of state, equation~(10).

\vskip 20pt
\noindent{\tenbf 7. Conclusions}

\nobreak

\vskip 5pt

\nobreak
We have described a new wave mode which is analogous to the acoustic wave, but involves oscillations of the tail of suprathermal distributions. These waves always travel faster than sound waves with the same temperature and in the limit as $\kappa\rightarrow\infty$ become standard acoustic waves.

The simple analysis presented here is based upon a one-dimensional collisionless neutral gas. In actuality, most physical contexts of interest are more complex, involving plasmas with electric and magnetic fields. However, the purpose of this paper is primarily to illustrate that investigators should consider the possibility
and look for evidence in the data of wave modes involving the suprathermal tail, that is $\kappa$. 

Recently, however, there has been some debate about low frequency (few Hz - kHz) broadband electrostatic emission representing ion acoustic wave activity versus solitary waves triggering a wide response in the waveform capture. One way to resolve this issue, at least at the lower frequencies below about 10~Hz, is to take in addition to electric and magnetic field measurements, complete ion distribution functions at frequencies close to 10~Hz or a cadence of 100~ms. Ion acoustic waves would be identified by correlations between density and velocity, In addition, possible non-Maxwellian generalizations of the ion acoustic wave, which are the subject of this paper, could be observed with such instrumentation.

Unfortunately, current magnetospheric particle instrumentation has never had a sufficient cadence to observe the new wave mode predicted in this paper.
One possibility is that the high-time resolution plasma investigations on the Magnetospheric Multiscale (MMS) mission currently scheduled to launch in 2014 may be able to detect these types of waves. The MMS/SMART Fast Plasma Instrument (FPI)
Dual Ion Spectrometers (DIS) will acquire complete distribution functions
up to 40~keV every 37.5 ms (four energy sweeps every 150~ms). The orbit during the first of the three MMS phases finds the four spacecraft formation in a 1.2x12~R${}_{\rm E}$ orbit at 28${}^\circ$ inclination which will permit high-time resolution studies of
both magnetospheric and solar wind plasmas.

\vskip 20pt
\noindent
{\tenbf Acknowledgments.} Thanks to Bill Farrell, John Podesta, and John Sigwarth for helpful conversations. 
\vfill \eject

\vskip 20pt
\noindent {\tenbf References}

\nobreak

\vskip 5pt

\nobreak

\hangindent=1.5pc \hangafter=1 \noindent
{Abramowitz, M. and I.A. Stegun, Handbook of Mathematical Functions with Formulas, Graphs, and Mathematical Tables, p. 258, 260, Dover Publications, Inc., New York, 1972.}

\hangindent=1.5pc \hangafter=1 \noindent
{Chen, Francis F., {\it Introduction to Plasma Physics and Controlled Fusion - Volume 1: Plasma Physics\/}, Plenum Press, New York, pp. 94-98, 1984.}

\hangindent=1.5pc \hangafter=1 \noindent
{Collier, Michael R., On generating kappa-like distribution functions using velocity space L\'evy flights, {\tenit Geophys. Res. Lett., 20\/}, 1531-1534, 1993.}

\hangindent=1.5pc \hangafter=1 \noindent
{Collier, Michael R., The adiabatic transport of superthermal distributions modeled by kappa functions, {\tenit Geophys. Res. Lett., 22\/}, 2673-2676, 1995.}

\hangindent=1.5pc \hangafter=1 \noindent
{Collier, Michael R., Evolution of kappa distributions under velocity space diffusion: a model for the observed relationship between their spectral parameters, {\tenit J. Geophys. Res., 104\/}, 28559-28564, 1999.}

\hangindent=1.5pc \hangafter=1 \noindent
{Collier, Michael R., Are magnetospheric suprathermal particle distributions ($\kappa$ functions) inconsistent with maximum entropy considerations?, {\tenit Adv. Space Res., 33\/}, 2108-2112, 2004.}

\hangindent=1.5pc \hangafter=1 \noindent
{Gurnett, D.A. and L.A. Frank, A region of intense plasma wave turbulence on auroral field lines, {\tenit J. Geophys. Res., 82\/}, 1031-1050, 1977.}

\hangindent=1.5pc \hangafter=1 \noindent
{Gurnett, D.A. and L.A. Frank, Plasma waves in the polar cusp: Observations from Hawkeye-1, {\tenit J. Geophys. Res., 83\/}, 1447-1462, 1978.}

\hangindent=1.5pc \hangafter=1 \noindent
{Gurnett, D.A., L.A. Frank, and R.P. Lepping, Plasma waves in the distant magnetotail, {\tenit J. Geophys. Res., 81\/}, 6059-6071, 1976.}

\hangindent=1.5pc \hangafter=1 \noindent
{Leubner, M.P., Fundamental issues on kappa-distributions in space plasmas and interplanetary proton distributions, {\tenit Phys. of Plasmas, 11(4)\/}, 1308-1316, 2004a.}

\hangindent=1.5pc \hangafter=1 \noindent
{Leubner, M.P., Core-halo distribution functions: A natural equilibrium state in generalized thermostatistics, {\tenit Astrophys. J., 604\/}, 469-478, 2004b.}

\hangindent=1.5pc \hangafter=1 \noindent
{Leubner, M.P. and Z. V\"or\"os, A nonextensive entropy approach to solar wind intermittency, {\tenit Astrophys. J., 618\/}, 547-555, 2005.}

\hangindent=1.5pc \hangafter=1 \noindent
{Samir, U., K.H. Wright, Jr., and N.H. Stone, The expansion of a plasma into a vacuum: Basic phenomena and processes and applications to space plasma physics, {\tenit Rev. Geophys. and Space Phys., 21\/}, 1631-1646, 1983.}

\hangindent=1.5pc \hangafter=1 \noindent
{Summers, D. and R.M. Thorne, The modified plasma dispersion function, {\tenit Phys. Fluids B, 3(8)\/}, 1835-1847, 1991.}

\hangindent=1.5pc \hangafter=1 \noindent
{Treumann, R.A., C.H. Jaroschek and M. Scholer, Stationary plasma states far from equilibrium, {\tenit Phys. of Plasmas, 11(4)\/}, 1317-1325, 2004.}

\hangindent=1.5pc \hangafter=1 \noindent
{Tsallis, Constantino, Non-extensive thermostatistics: brief review and comments, {\tenit Physica A, 221\/}, 277-290, 1995.}

\vfill \eject

\noindent {\tenbf Figure Captions}

\nobreak

\vskip 5pt

\nobreak

\noindent
{\tenbf Figure 1.} The one and three-dimensional $\xi$ functions which modify the standard Maxwellian expressions for conservation of entropy.
\vskip 0.10in

\noindent
{\tenbf Figure 2.} The ratio of the phase velocity of the one-dimensional acoustic kappa wave normalized by the phase speed of a standard acoustic wave with the same temperature.
\vskip 0.10in

\vfill \eject
\end